\documentclass[11pt]{article}

\usepackage[top=0.7in, bottom=0.7in, left=1.2in, right=1.2in]{geometry}

\usepackage{amsmath, amssymb, graphicx}
\usepackage{hyperref}
\usepackage{graphicx}
\usepackage{media9}
\usepackage{float}
\title{Unveiling the Dynamics of Employee Behavior Through Wolfram’s Cellular Automata}
\author{Rakshitha Jayashankar, Mahesh Balan}
\date{}

\begin{document}

\maketitle

\begin{abstract}
 Understanding employee behavior in a workplace is critical for enhancing overall organizational performance. Despite numerous efforts to improve work environments, many organizations still need help with challenges primarily rooted in unaddressed issues or poorly understood behavioral patterns. In this paper, we have focused on recognizing this pattern and the dynamics of complex systems in organizational behavior and studying how factors influence the system’s overall behavior using Wolfram’s Cellular Automata theory. Over the cycle, we observe how leadership influences team dynamics, influences the organization, and drives employee behavior to foster a positive environment.
\textbf{Keywords}: Cellular Automata, Employee Behavior, team dynamics, Leadership Influence
\end{abstract}

\section{Introduction}
Employee behavior in the workplace is a critical factor that influences the organization's culture, productivity, and success. Active and energetic behaviors are useful for organizations to obtain their goals and vice versa \cite{1}. Organizations can utilize various methods to predict employee behavior, such as surveys and questionnaires, observational studies, interviews and focus groups, behavioral analytics, psychometric testing, and social network analysis. It requires a framework that can capture the intricacies of interactions among individuals and the emergence of patterns over time. In this quest, Wolfram's Cellular Automata provides insights into the dynamics of complex systems. This paper delves into the application of Cellular Automata to unravel the mechanisms driving employee behavior and foster a positive work environment.

Agent-based modeling, a key feature of Cellular Automata, is a powerful tool that effectively captures the dynamics of very complex systems and their dynamic interactions. It does so by simulating emergent behavior from simple rules, providing a unique lens to understand employee behavior in the workplace.

\section{Understanding the working of Employee Behavior using 2D cellular automata}
John von Neumann, a pioneer in the field, originally proposed cellular automata as formal models of self-reproducing organisms. His work, which focused on one- and two-dimensional infinite grids, laid the foundation for our understanding of these structures, though higher dimensions were also considered. This seminal work continues to inspire and guide our research today \cite{2}.

Cellular automata, a subject of continuous exploration, are being studied in numerous fields today. The 'Game of Life,' introduced by British mathematician John Conway, is a prime example. This 2D cellular automaton with simple rules that can exhibit complex behavior has been instrumental in our understanding of the potential of cellular automata. Stephen Wolfram's extensive research on cellular automata, classifying their behaviors and dynamics, is a testament to the evolving nature of this field. Wolfram's research further provided a systematic study and classification of CA for computing complex behavior through discrete rules.

Using 2D CA can provide very valuable insights and benefits for understanding and managing complex social dynamics in an organization. Applying cellular automata in organizational dynamics can empower upper management by providing them with the tools to make informed decisions, foster a positive work environment, and effectively achieve strategic objectives. This practical application of our model underscores its relevance and potential impact in real-world scenarios.

Our model is firmly grounded in the tangible application of Wolfram's Theory, specifically a 2D cellular automaton. This automaton serves as the bedrock of our model, with each cell representing an employee. The evolution of these cells over discrete time steps, guided by predefined rules based on their current state and neighbors' states, directly mirrors our research on employee behavior.

\section{Methodology}
\subsection{Model Simulation}
\subsubsection{Initializing the Grid}
The model, which utilizes cellular automata to simulate team dynamics and Leadership Influence, is a significant tool in understanding the complexity of interaction and gaining valuable insights for organizational management. This involves initializing the simulation environment, defining rules for updating employee performance states, running all the simulations, and visualizing the result.

The 20*20 grid, a crucial component of our simulation, represents each employee's performance state. A random performance state level ranging from 0 to 5 (lowest to highest) is assigned to each grid cell. This random assignment and initialization of the grid is the starting point that exists in real-world scenarios.

The Grid Representation $G(t)$ is a crucial element in our model where $G_{ij}(t)$ represents the performance state of the employee at position $(i,j)$ on the grid. This performance state $G_{ij}(t)$ is an integer value in the range [0, 5]. The grid's dynamic nature, as captured by $G(t)$, allows for a comprehensive view of employee performance states over time, facilitating a deeper analysis of the simulation results.

\subsubsection{Simulation Cycles}
We meticulously define the number of simulations (Cycle simulation) from 1 to 256. Each cycle represents a discrete time step on which each employee's performance is measured and updated on the next cycle based on interaction with the neighboring employee. This precision allows us to measure and analyze the grid's evolution and the grid's average performance state at each step.

Our model operates iteratively. A new grid is created for each calculation, mirroring the current grid but with updated performance states. Each grid is defined within the grid boundaries. The performance state in each cycle is the average value based on the neighboring cells within the grid. This iterative process, where the average of the surrounding cells determines the influence on the central cell, allows for continuously updating performance states.

\subsubsection{Average Neighbor State}
The average performance state of the neighbors of $G_{ij}(t)$ is calculated as follows:
\[
\text{avg\_neighbor}(i, j, t) = \frac{1}{|N_{ij}|} \sum_{(k,l) \in N_{ij}} G_{kl}(t)
\]

Where $N_{ij}$ is the number of valid neighboring cells. The Evolution of Employee Behavior depicts the average performance of the cycle from 0 to 255.

\begin{figure}[H] 
  \centering
  \includegraphics[width=0.85\textwidth]{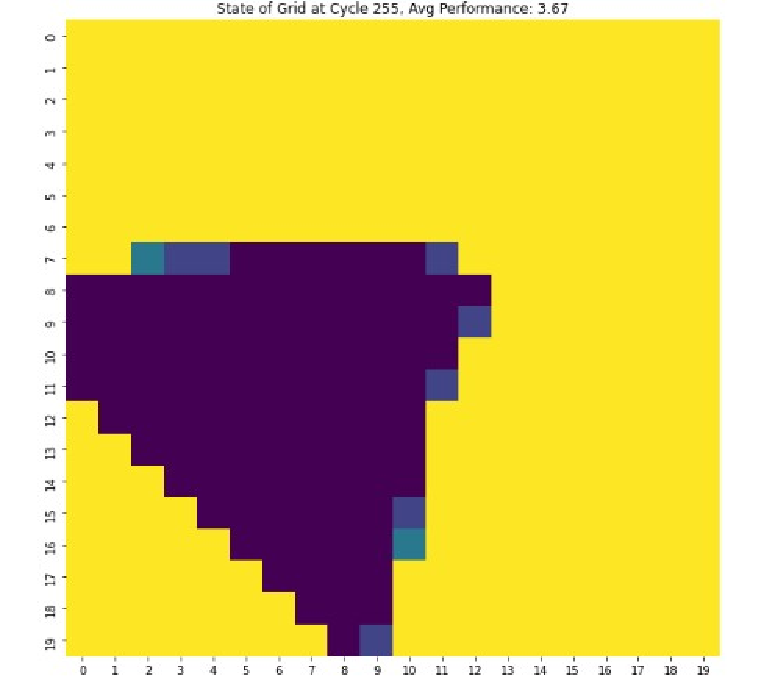}
  \caption{State of the cells when Average performance 3.67}
\end{figure}
Link to the animation simulating all states:
\href{https://github.com/rakshithajayashankar/EmployeeBehavior-using-Cellular-Automata/blob/main/animation.gif}{GitHub Repository}

\subsection{Factors Affecting the Performance of the Employee}
\subsubsection{Leadership Influence}
Leadership influence is a critical factor in shaping the behavior of the employees at the workplace. The rules that adjust the performance state are defined in the simulation model. Effective leadership can inspire higher productivity and performance, improve morale, and foster a more motivated and better functional work environment. The leadership model, with its key assumption that high-performing neighbors represent strong leadership, empowers and motivates the performance of employees around them.

\begin{figure}[H] 
  \centering
  \includegraphics[width=0.92\textwidth]{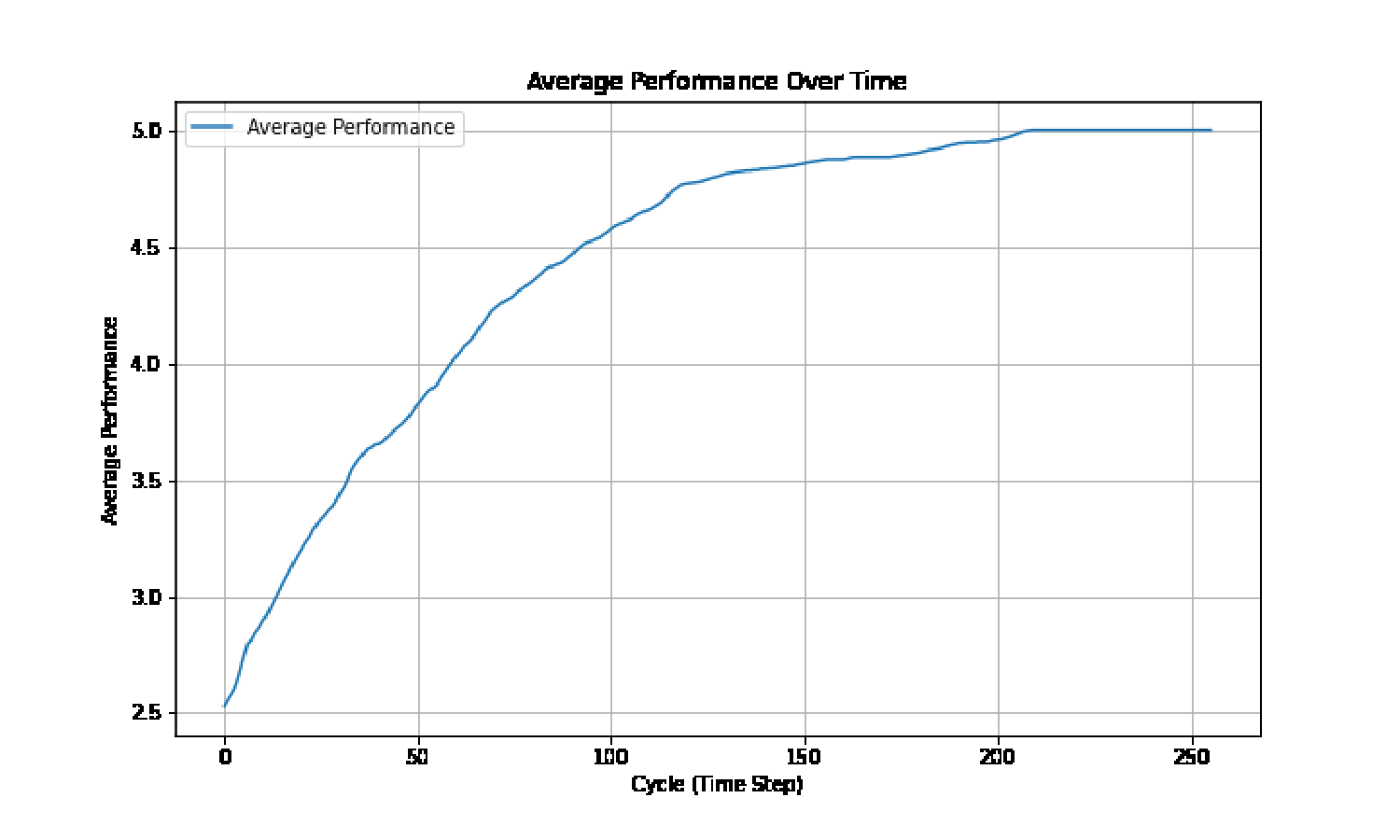}
  \caption{Average Performance over the cycles}
\end{figure}

The essence of Leadership Influence lies in recognition of effective leadership. It underscores the fact that high-performance neighbors, often perceived as leaders, serve as the wellsprings of positive influence, fostering a positive culture.

If the average performance in each cell of the neighbors exceeds 3, it signifies a potent leadership influence. This leads to a 1-point increase in the performance of the central cell, up to a maximum value of 5, demonstrating the power of strong leadership influence.

In this simulation, a performance state of over 3 for the average of neighbors is used as a heuristic approach. It symbolizes an environment where leadership influence is robust enough to significantly boost an employee’s performance, thereby fostering a sense of optimism in the potential outcomes.

Reiterating the concept of effective leadership, we see high-performing neighbors as reliable sources of positive influence in the organization. By modeling the threshold, we demonstrate that strong leadership or a high-performing team can elevate the performance of individuals in their proximity. The threshold value of 3 is used as a heuristic to define high leadership influence because it represents a level of performance above the average. This rule helps to simulate the positive impact of being surrounded by high-performing peers, which can be seen as analogous to strong leadership influence in an organizational context. This reiteration provides reassurance about the reliability of this model, instilling confidence in its effectiveness.

\subsubsection{Team Dynamics}
The cellular automata framework, a significant tool in our research, is instrumental in understanding the team dynamics that influence collective performance. It reveals that a positive influence from the team can lead to performance improvements. In contrast, a negative impact from the team can result in low-performing neighbors, potentially leading to a performance decline.

In this paper, the implementation of team dynamics discusses the mechanism of all the positive influences on neighboring cells.

The positive influence of high-performing neighbors (cells with a performance state above 3) is a beacon of optimism. Their contribution elevates a cell's performance state, showcasing the advantageous effect of collaborating with high-performing team members or operating under effective leadership.
\[ 
G_{ij}(t + 1) = \min(G_{ij}(t) + 1, 5) \quad \text{if} \quad \text{avg\_neighbor}(i, j, t) > 3 
\]

\begin{figure}[H] 
  \centering
  \includegraphics[width=0.92\textwidth]{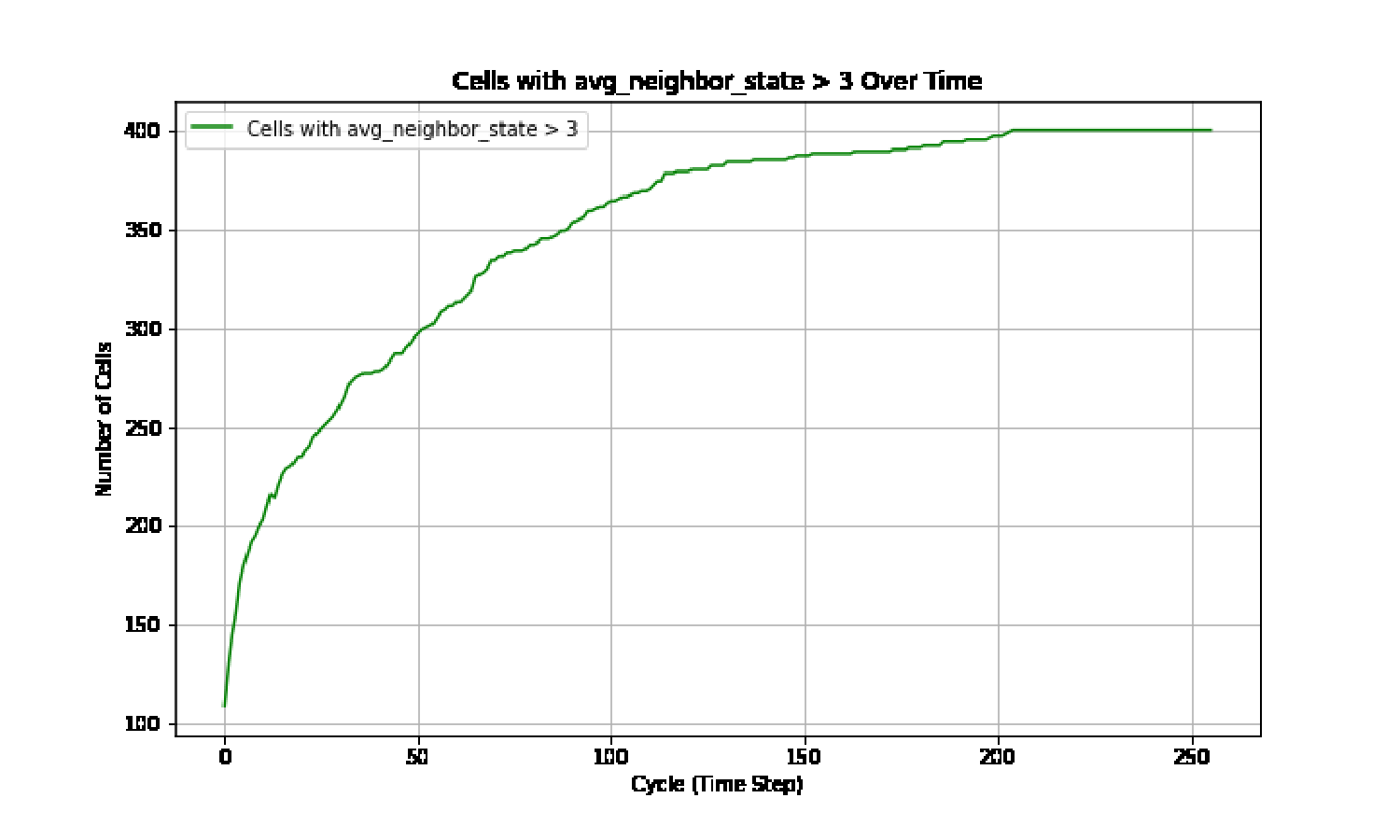}
  \caption{Average performance above 3}
\end{figure}

The Negative Influence manifests when low-performing neighbors (cells with a performance state below 2) decrease a cell's performance state.
\[ 
G_{ij}(t + 1) = \max(G_{ij}(t) - 1, 0) \quad \text{if} \quad \text{avg\_neighbor}(i, j, t) < 2 
\]

\begin{figure}[H] 
  \centering
  \includegraphics[width=0.92\textwidth]{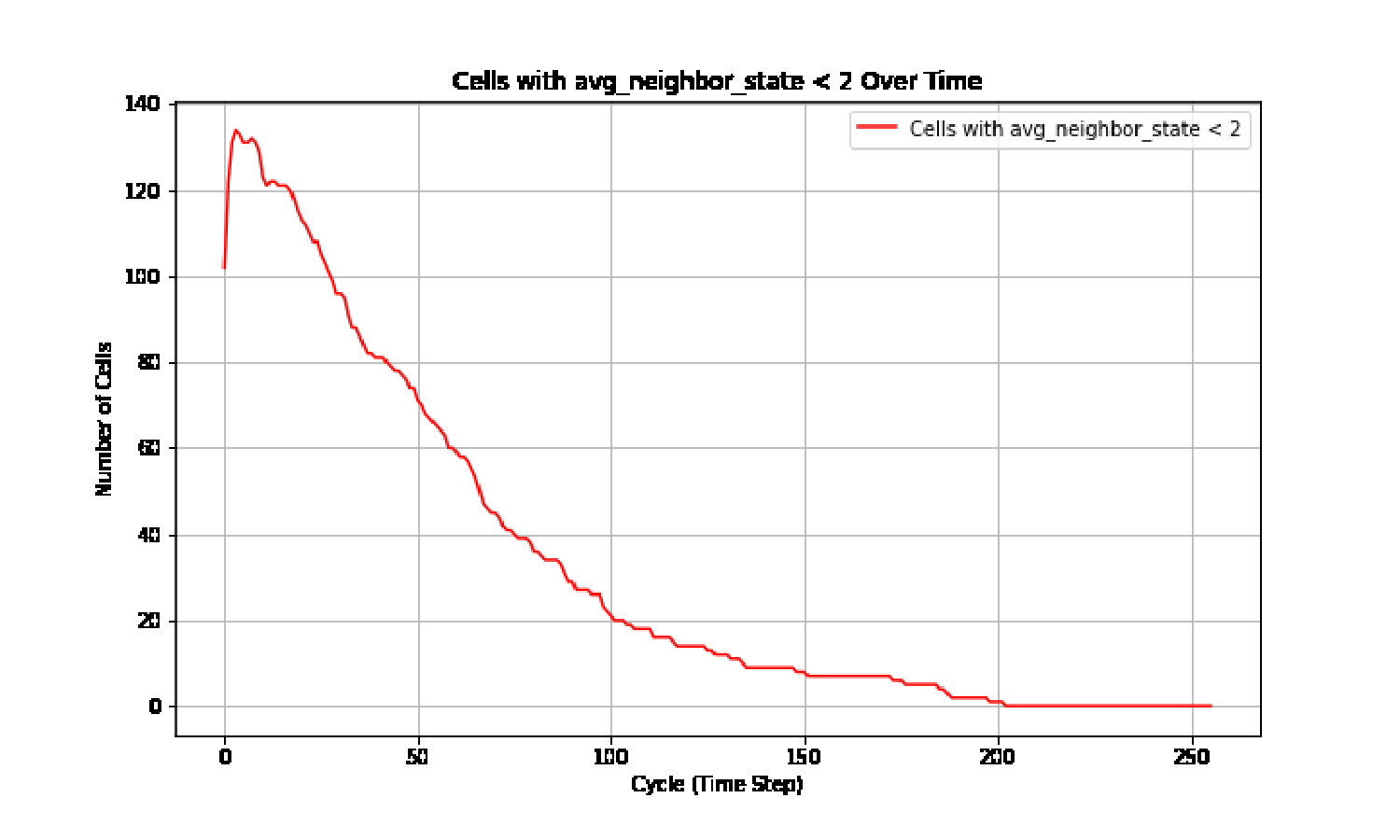}
  \caption{Average performance Below 2}
\end{figure}

This mirrors the detrimental effect of working with underperforming team members, a factor that warrants careful consideration.
\[
G_{ij}(t + 1) = G_{ij}(t) \quad \text{if} \quad \text{avg\_neighbor}(i, j, t) = 3
\]

\begin{figure}[H] 
  \centering
  \includegraphics[width=0.92\textwidth]{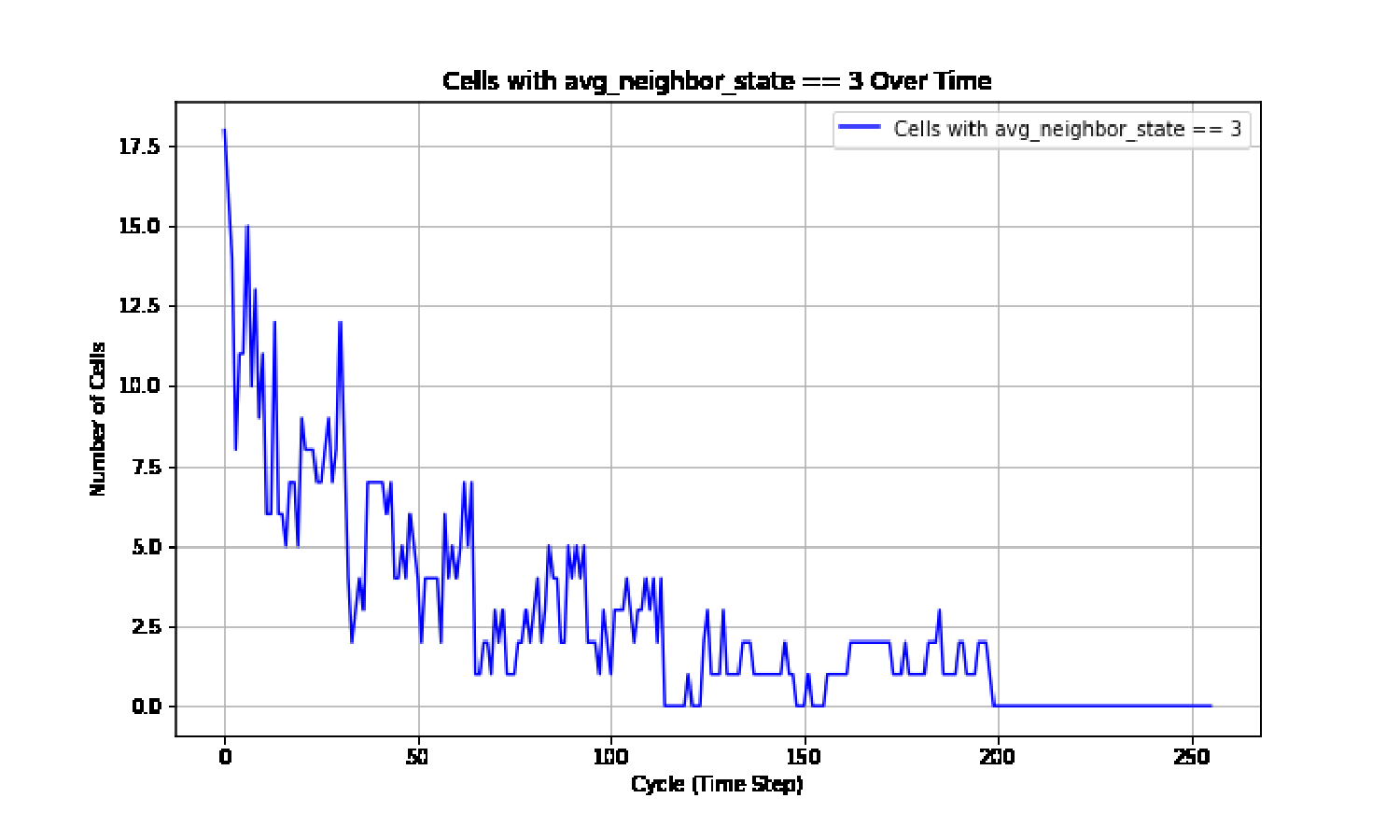}
  \caption{Average performance equal to 3}
\end{figure}

\section{Conclusion}
The dynamics and application of cellular automata, a computational model that operates on a grid of cells in organizational behavior and how leadership and team dynamics help us understand the complex behavior of employees and interactions within the organizations. For instance, this paper demonstrates how Cellular Automata can simulate and analyze the effects of leadership influence and team dynamics on employee performance over time.

\subsection{Key Findings for the simulation}
\begin{itemize}
    \item Positive Leadership Influence: The cellular automata model vividly demonstrates the pivotal role of high-performing employees in positively influencing their neighboring employees when their average performance exceeds a threshold. This mirrors the real-world scenario in organizations where good leadership and high-performing individuals inspire and elevate performance. This finding underscores the significant influence that leaders like you can have on team dynamics and overall performance.
    \item Team Dynamics: The simulation underscores the crucial role of interactions between employees in the workplace. Modeled by the performance states of neighboring cells, these interactions significantly impact overall team performance. Negative dynamics, where the average performance of neighbors is low, result in a decrease in an individual's performance state. This finding highlights the urgency of maintaining a supportive and high-performing team environment. The model also incorporates random rewards and sustained performance boosts, which simulate real-world incentives and recognition systems that can further enhance performance.
\end{itemize}

By using cellular automata to simulate these interactions, this study not only provides a quantitative framework to explore the impact of various factors on employee behavior but also offers practical insights for managers and organizational researchers. It underscores the potential of CA as a tool to better understand and optimize team dynamics and leadership strategies, thereby enhancing the performance and culture of the organization. Future research could further enhance this model by incorporating additional variables and more complex rules, thereby increasing its applicability and accuracy.

In conclusion, cellular automata offer a robust and flexible method for modeling the intricate patterns of team dynamics and leadership influence in organizational behavior. This approach enhances our theoretical understanding and provides practical insights for improving employee performance and fostering a positive organizational culture.

Here is a link to my GitHub repository: \href{https://github.com/rakshithajayashankar/EmployeeBehavior-using-Cellular-Automata/blob/main/Employee%20Behavior_Leadership%20Influence%20and%20Team%20Dynamics.ipynb}{GitHub Repository}


\begin{thebibliography}{9}
\bibitem{1} Jiao Y. Sun S. Sun X. (2007). Simulation of Employee Behavior Based on Cellular Automata Model. In: Shi Y. van Albada G.D. Dongarra J. Sloot P.M.A. (eds) Computational Science – ICCS 2007. ICCS 2007. Lecture Notes in Computer Science, vol 4490. Springer, Berlin, Heidelberg. \url{https://doi.org/10.1007/978-3-540-72590-9_18}
\bibitem{2} Palash Sarkar. 2000. A brief history of cellular automata. ACM Comput. Surv. 32, 1 (March 2000), 80–107. \url{https://doi.org/10.1145/349194.349202}
\bibitem{3} Somers M.J. Ethical Codes of Conduct and Organizational Context: A Study of the Relationship Between Codes of Conduct, Employee Behavior, and Organizational Values. Journal of Business Ethics, 30, 185–195 (2001). \url{https://doi.org/10.1023/A:1006457810654}
\bibitem{4} Braz L.F. Sichman J.S. (2022). Using MBTI Agents to Simulate Human Behavior in a Work Context. In: Czupryna M. Kamiński B. (eds) Advances in Social Simulation. Springer Proceedings in Complexity. Springer, Cham. \url{https://doi.org/10.1007/978-3-030-92843-8_25}
\bibitem{5} E. Weisstein “Elementary Cellular Automaton” MathWorld, Wolfram Research. \url{https://mathworld.wolfram.com/ElementaryCellularAutomaton.html}
\bibitem{6} E. Weisstein “Game of Life” MathWorld, Wolfram Research. \url{https://mathworld.wolfram.com/GameofLife.html}
\end{thebibliography}
\end{document}